\documentstyle[11pt]{article}
\begin{document}

\begin{center}
{\bf EXPLICIT SUMMATION OF THE CONSTITUENT WKB SERIES AND NEW
APPROXIMATE WAVE FUNCTIONS }\\
 \medskip
{ VLADIMIR V. KUDRYASHOV\dag\ and YULIAN V. VANNE\ddag }\\
\medskip
{\small\it \dag\ Institute of Physics, National Academy of
 Sciences of Belarus,\\
 68 F. Skaryna Avenue , 220072 Minsk, Belarus \\
\ddag\ Department of Chemistry, University of Konstanz,\\
 Fach M721 D-78457, Konstanz, Germany \\}

{\small\it e-mail: kudryash@dragon.bas-net.by and
yulian.2.vanne@uni-konstanz.de}
\end{center}
\begin{abstract}
The independent solutions of the one-dimensional Schr\"odinger
equation are approximated by means of the explicit summation of
the leading constituent WKB series. The continuous matching of
the particular solutions gives the uniformly valid analytical
approximation to the wave functions. A detailed numerical
verification of the proposed approximation is performed for some
exactly solvable problems arising from different kinds of
potentials.
\end{abstract}
\medskip
2000 Mathematics Subject Classification 81Q05, 81Q20, 34E05, 40A30

\section{Introduction}
 Perturbation theory, the variational method and the WKB
approximation are very extensively used in quantum mechanics.
   If we deal with perturbation theory or with the variational method
then similar questions arise. How to find the unperturbed
Hamiltonian or how to find the trial function for an arbitrarily
given potential? Universal answers are absent. In this sense both
mentioned methods are incomplete. In contrast, the WKB
approximation is directly determined by a given potential.
However the conventional WKB approximation has unphysical
singularities. An old problem in semiclassical analysis is the
development of global uniform approximations to the wave
functions. In previous works \cite{kudr1,kudr2}, an essential
improvement of the WKB approach was introduced for the logarithmic
derivatives of the wave functions. In the present paper, we
construct  the second-order continuous approximation to the wave
functions. The quality of the approximate wave functions is
verified by means of a comparison with the exact solutions for
different kinds of potentials.

   We consider the linear one-dimensional Schr\"odinger equation
\begin{equation}
\frac{d^2 \Psi(q,\hbar)}{d q^2} =  \frac{Q(q)}{\hbar^2}
\Psi(q,\hbar)  ,
\end{equation}
where $ Q(q) = 2m \left( V(q) - E \right) $ for an arbitrary
potential $ V(q) $. The logarithmic derivative \[ Y(q,\hbar)
=\frac{d \ln \Psi(q,\hbar)}{dq} \] of a wave function
$\Psi(q,\hbar)$ satisfies the nonlinear Riccati equation
\begin{equation}
\label{Riccati}
 \frac{dY(q,\hbar)}{dq} + \left(Y(q,\hbar)\right)^2 =
\frac{Q(q)}{\hbar^2} .
\end{equation}
The WKB approach deals just with functions $Y(q,\hbar)$. In this
approach, two independent solutions $ Y^{\pm}(q,\hbar) $ of the
Riccati equation are represented by their asymptotic expansions
\begin{equation}
\label{wkb}
 Y^{\pm}_{as}(q,\hbar) =  \hbar^{-1} \left(\pm  Q^{1/2}
+
 \sum_{n = 1}^{\infty}\hbar^n Y^{\pm}_n (q) \right)
\end{equation}
in powers of Plank's constant $\hbar$. The usual WKB
approximation contains a finite number of leading terms
$Y_n^{\pm}(q)$ from the complete expansions $
Y^{\pm}_{as}(q,\hbar)$. This approximation is not valid at turning
points where $Q(q) = 0$.

 As it is well known, the WKB series is divergent. Numerous
 references regarding  asymptotic expansions may be found in
 \cite{boyd}. The direct summation of  a divergent series does not exist.
By {\it summing} one means finding a function to which this series
is the asymptotic expansion \cite{bend1}. In recent years many
studies have been devoted to extracting some useful information
about the exact eigenfunctions from the divergent WKB series
(see, e.g., \cite{pham} and references therein). There are several
investigations on the properties of the WKB terms
\cite{bend2,robn}. Unlike entirely exact but very complicated
methods for some classes of potentials (see, e.g., \cite{voros})
our new way of using the WKB series gives an approximate but very
simple and universal method of solving the Schr\"odinger equation.

\section{Explicit summation of the constituent WKB series}

Since \cite{kudr2} is likely to be inaccessible for the large
majority of readers we reproduce previous results. First of all,
the analysis of the well-known recursion relations
\cite{bend1,bend2}
\begin{equation}
\label{recur}
 Y^{\pm}_{n+1} = -(2Y^{\pm}_{0})^{-1}\left( \sum^{n}_{j=1}
 Y^{\pm}_{j}Y^{\pm}_{n+1-j}
 + dY^{\pm}_{n}/dq \right), \quad Y^{\pm}_0 =\pm Q^{1/2}
\end{equation}
shows that the WKB terms are of the form
\begin{equation}
\label{rec}
 Y^{\pm}_{n}(q) = Q^{\frac{1-3n}{2}} \sum^{n}_{j=1}A^{\pm}_{n,j}\bigl(
 Q'(q),Q''(q),...Q^{(j)}(q) \bigr)Q^{j-1}  ,
\end{equation}
where  $ Q'(q) = d Q(q)/d q$, $Q''(q) = d^2 Q(q)/d q^2$ and
$Q^{(j)}(q) = d^j Q(q)/d q^j$.  Second, the substitution of
(\ref{rec}) into (\ref{wkb}) allows us to reconstruct the
asymptotic WKB series  as an infinite sum
 \[
 Y^{\pm}_{as}(q,\hbar)
= \pm \hbar^{-1} Q^{1/2}
 + \sum^{\infty}_{j = 1} Z^{\pm}_{as,j}(q,\hbar)
\]
 of new constituent (partial) asymptotic series
\[
 Z^{\pm}_{as,j}(q,\hbar) =
 (\hbar^{2/3})^{j - 2} \sum^{\infty}_{n = j}
 \left( \frac{Q}{\hbar^{2/3}} \right)
^{j - \frac{3n + 1}{2}} A^{\pm}_{n,j}(Q'(q),Q''(q),...Q^{(j)}(q))
\]
 in powers of the ratio $Q/\hbar^{2/3}$. With the help of the
recursion relations (\ref{recur})  we derive  simple expressions
\[
 A^{\pm}_{n,1} = (Q')^n  B^{\pm}_{n,1} ,\quad
 A^{\pm}_{n,2} =Q''(Q')^{n - 2} B^{\pm}_{n,2}
\]
 for two leading sequences of coefficients $A^{\pm}_{n,j}$ .
Here the numbers $B^{\pm}_{n,1}$ are determined by the following
recursion relations:
\begin{equation}
\label{B1}
 B^{\pm}_{n+1,1} = {\mp} \left(\frac{1}{2}
\sum^n_{k=1}B^{\pm}_{k,1} B^{\pm}_{n+1-k,1} +\frac{1 - 3 n}{4}
B^{\pm}_{n,1} \right), \quad n\geq1, \quad B^{\pm}_{1,1} =
-\frac{1}{4}  ,
\end{equation}
 and $B^{\pm}_{n,2}$ is connected with
$B^{\pm}_{n,1}$ as follows:
\begin{equation}
\label{B2}
 B^{\pm}_{n,2} = -\frac{2}{5} n
B^{\pm}_{n,1}  , \quad  n\geq2.
\end{equation}

   The complete series $Y^{\pm}_{as}(q,\hbar)$ are approximated by
a finite number of leading constituent series
$Z^{\pm}_{as,j}(q,\hbar)$ in contrast to the use of a finite
number of leading  terms $Y^{\pm}_n(q)$ in the conventional WKB
approach. If we can find functions $Z^{\pm}_j(q,\hbar)$ which are
represented by asymptotic expansions $Z^{\pm}_{as,j}(q,\hbar)$,
then we obtain new approximations to the solutions of the Riccati
equation. The number of used constituent series corresponds to the
order of a proposed approximation. For instance the expressions
$\pm\hbar^{-1} Q^{1/2} + Z^{\pm}_1(q,\hbar)$ are interpreted as
the first-order approximations. In this paper, we consider only
the second-order approximations  $\pm\hbar^{-1} Q^{1/2} +
Z^{\pm}_1(q,\hbar) + Z^{\pm}_2(q,\hbar)$.

    Introducing the dimensionless variable
\begin{equation}
a(q,\hbar) = \frac{1}{\hbar^{2/3}} \frac{Q(q)}{|Q'(q)|^{2/3}}  ,
\end{equation}
we are able to rewrite the leading constituent expansions in the
form
\begin{eqnarray}
 \pm \hbar^{-1} Q^{1/2} + Z^{\pm}_{as,1}(q,\hbar) +
Z^{\pm}_{as,2}(q,\hbar)
 =  \frac{1}{\hbar^{2/3}} \frac{Q'}{|Q'|^{2/3}} y^{\pm}_{as,1}(a) +
 \frac{Q''}{Q'} y^{\pm}_{as,2}(a)  ,
\end{eqnarray}
where we separate the asymptotic series in $a$
\begin{equation}
\label{yas1}
 y^{\pm}_{as,1}(a) = \pm a^{1/2} +   \sum^{\infty}_{n = 1}
B^{\pm}_{n,1} a^{1 - \frac{3n + 1}{2}} ,
\end{equation}
\begin{equation}
\label{yas2}
 y^{\pm}_{as,2}(a) = \sum^{\infty}_{n = 2} B^{\pm}_{n,2} a^{2-
\frac{3n + 1}{2}} .
\end{equation}
The leading terms
\begin{equation}
\label{ym1}
  \pm a^{1/2} - \frac{1}{4} a^{-1} ,
\end{equation}
\begin{equation}
\label{ym2}
  \pm \frac{1}{8} a^{-3/2} + \frac{9}{32} a^{-3}
\end{equation}
of these series may be deduced by using equations (\ref{B1}) and
(\ref{B2}).

   Our aim is to sum constituent series (\ref{yas1}) -- (\ref{yas2}). In other
words, we must find functions $y^{\pm}_j(a)$ which are
represented by these expansions. In order to perform the
identification we substitute the approximate function
\begin{equation}
\label{Ypm_ap}
 Y^{\pm}_{ap}(q,\hbar) =
 \frac{1}{\hbar^{2/3}} \frac{Q'}{|Q'|^{2/3}} y^{\pm}_1(a)
+ \frac{Q''}{Q'} y^{\pm}_2(a)
\end{equation}
into the Riccati equation (\ref{Riccati}). As a result we get the
following equations:
\begin{equation}
\label{dy1_a}
 \frac{dy^{\pm}_1}{da} + (y^{\pm}_1)^2 = a ,
\end{equation}
\begin{equation}
\label{dy2_a}
 \frac{dy^{\pm}_2}{da} + 2y^{\pm}_1y^{\pm}_2 = \frac{1}{3}\left( 2a\frac{dy^{\pm}_1}{da} -
y^{\pm}_1\right)
\end{equation}
for the functions $y^{\pm}_j(a)$. Direct verification shows that
the asymptotic expansions (\ref{yas1}) and (\ref{yas2}) satisfy
these equations.

    Equation (\ref{dy1_a}) is the Riccati equation for  the logarithmic
derivatives of linear combinations of the well-studied Airy
functions ${\rm Ai}(a)$ and ${\rm Bi}(a)$ \cite{hand}. We select
particular solutions by means of the known asymptotics
(\ref{ym1}). In the classically allowed region where $Q(q) < 0~
(a < 0) $ we derive the explicit expressions
\begin{equation}
\label{y_pm1}
 y^{\pm}_1(a) = \frac{d}{da} \ln\left({\rm Bi}(a) \mp i{\rm Ai}(a)
\right)
\end{equation}
and in the classically forbidden region where $Q(q) > 0~ (a > 0)$
we get the other solutions
\begin{equation}
\label{tildey_pm1} \tilde y^-_1(a) = \frac{d}{da} \ln {\rm Ai}(a) ,\quad \tilde
y^+_1(a) = \frac{d}{da} \ln {\rm Bi}(a) .
\end{equation}

   Finally, we can obtain the solutions of the linear equation (\ref{dy2_a}) with
asymptotics (\ref{ym2}) in the closed form
\begin{equation}
\label{y_pm2}
 y^{\pm}_2(a) =  \frac{1}{30} \left[-8 a^2 (y^{\pm}_1(a))^2  -4 a
y^{\pm}_1(a) +
 8 a^3 - 3 \right] ,
\end{equation}
\begin{equation}
\label{tildey_pm2}
 \tilde y^{\pm}_2(a) =  \frac{1}{30} \left[-8 a^2 (\tilde
y^{\pm}_1(a))^2  -4 a \tilde y^{\pm}_1(a) +
 8 a^3 - 3 \right] .
\end{equation}

Although the functions (\ref{y_pm1}) -- (\ref{tildey_pm2}) have
the asymptotic expansions (\ref{yas1}) -- (\ref{yas2}) if $|a|$ is
large it should be stressed that the obtained functions possess
different expansions if $|a|$ is small. Replacing $y_j^{\pm}$ by
$\tilde y_j^{\pm}$ in expression (\ref{Ypm_ap}) we get the second
pair
\begin{equation}
\tilde Y^{\pm}_{ap}(q,\hbar) =
 \frac{1}{\hbar^{2/3}} \frac{Q'}{|Q'|^{2/3}} \tilde y^{\pm}_1(a)
+ \frac{Q''}{Q'} \tilde y^{\pm}_2(a)
\end{equation}
of approximate solutions.

   It is not surprising that the asymptotics of our approximation
coincide with the WKB asymptotics far away from  the turning
points. At the same time our approximation reproduces the known
\cite{bend1} satisfactory approximation near  the turning points.
Naturally, our approximation gives the exact result for the
linear potential $V(q) = k q$ of a uniform field. Note that this
potential represents an example of  the explicit summation of the
WKB series for the logarithmic derivative of a wave function.

\section{Approximate wave functions for the two-turning-point problem}
   With the aid the uniformly valid approximation to solutions of
the Riccati equation derived in preceding section, we can now
construct approximate wave functions. We consider the problem
with two real turning points $q_-$ and $q_+$ $(q_+ > q_-)$. The
potential has its minimum at point $q_m$. The first and second
derivatives of the smooth potential are continuous at point $q_m$.

    Two pairs of independent solutions of the Schr\"odinger equation
are approximated by functions
\[
\Psi_{ap}^{\pm}(q) = \exp\ \left(\int^q Y_{ap}^{\pm}(q') d
q'\right)  ,
\]
\[
\tilde \Psi_{ap}^{\pm}(q) = \exp \left(\int^q
 \tilde Y_{ap}^{\pm}(q') d q'\right) .
\]
In accordance with the requirements of quantum mechanics, we must
retain only the decreasing solutions $\tilde \Psi_{ap}^-(q)$ in
the classically forbidden regions ($q < q_-$ and $q > q_+$). In
the classically allowed region ($q_- < q <q_+$) we retain a linear
combination of two oscillatory  solutions $\Psi_{ap}^{+}(q)$ and
$\Psi_{ap}^{-}(q)$. By matching particular solutions at the
turning points  $q_-$ and $q_+$, we obtain the continuous
approximate wave function which is represented by following
formulas:
\begin{equation}
\Psi_1(q) = C \cos{\frac{\pi}{3}}
 \exp\left( -\int_q^{q_-} \tilde Y_{ap}^-(q') dq' \right)
\end{equation}
if $q < q_-$,
\begin{eqnarray}
 \Psi_2(q) = C \exp \left( \int _{q_-}^q
 \frac{Y_{ap}^+(q') + Y_{ap}^-(q')}{2} d q' \right)
 \cos \left(\int_{q_-}^q \frac{Q'}{|Q'|}
 \frac{Y_{ap}^+(q') - Y_{ap}^-(q')}{2 i} d q' -
\frac{\pi}{3} \right)
\end{eqnarray}
if $q_- < q < q_+$, and
\begin{eqnarray}
 \Psi_3(q) = C (-1)^n \cos{\frac{\pi}{3}}
 \exp\left( \int_{q_+}^q \tilde Y_{ap}^-(q') d q' \right)
  \exp\left( \int_{q_-}^{q_+} \frac{Y_{ap}^+(q') +
 Y_{ap}^-(q')}{2} d q'\right)
\end{eqnarray}
if $q > q_+$.

  Here we have the new quantization condition
\begin{equation}
\int_{q_-}^{q_+} \frac{Q'}{|Q'|} \frac{Y_{ap}^+(q,E) -
 Y_{ap}^-(q,E)}{2 i} d q
= \pi (n + \frac{2}{3}) , \qquad n = 0,1,2,...
\end{equation}
which determines the spectral value  $E_{sp}(n)$ of energy
implicitly. We denote the wave functions with $E = E_{sp}(n)$ as
$\Psi_{ap}(q,n)$. Then we may choose  the value of an arbitrary
constant $C$ in order to ensure the usual normalization $\langle
\Psi_{ap}(n)|\Psi_{ap}(n)\rangle = 1$ where
$|\Psi_{ap}(n)\rangle$ is the vector in Hilbert space which
corresponds to the function $\Psi_{ap}(q,n)$. The proposed
approximation is an alternative to the well-known
\cite{bend1,olver} Langer approximation \cite{lang} which employs
a $\hbar$-expansion different from the WKB series.

Thus the approximate eigenfunctions are determined completely.
However a question arises regarding the optimal approximate
eigenvalues, because the value $E_{sp}(n)$ is not a unique choice.

  Since explicit expressions for wave functions have already been
  obtained,
we are able to calculate the expectation values
\begin{equation}
\bar E(n) = \langle \Psi_{ap}(n)|\hat H|\Psi_{ap}(n)\rangle
\end{equation}
of the Hamiltonian
\[
\hat H = -\frac{\hbar^2}{2 m}\frac{d^2}{d q^2} + V(q) .
\]
In accordance with the eigenvalue problem
\[
\hat H |\Psi\rangle -E |\Psi\rangle = 0
\]
we construct the discrepancy vector
\[
|D(e,n)\rangle = \hat H |\Psi_{ap}(n)\rangle - e ,
|\Psi_{ap}(n)\rangle  ,
\]
where $e$ is an arbitrary parameter while $\hat H$ and
$|\Psi_{ap}(n)\rangle$ are given. It is natural to require that
the discrepancy vector should not contain a component
proportional to the approximate eigenvector. In other words, we
consider the orthogonality condition
\[
\langle \Psi_{ap}(n)|D(e,n)\rangle = \bar E(n) - e = 0
\]
as a criterion for the selection of the optimal approximate
eigenvalue. As a result we just get $\bar E(n)$ while $E_{sp}(n)$
does not fulfil the above requirement. It should also be noted
that the scalar product $\langle D(e,n)|D(e,n) \rangle$  is
minimized at $e=\bar{E}(n)$.

\section{Verification of the proposed approximation}

   Now we must verify our approximation numerically for exactly
solvable problems. We compare the normalized approximate wave
functions $\Psi_{ap}(q,n)$ with the normalized exact wave
functions $\Psi_{ex}(q,n)$.

   In order to estimate the closeness of two functions $f_1(q)$
and $f_2(q)$, we consider two corresponding vectors $|f_1\rangle$
and $|f_2\rangle$ in Hilbert space. Then we construct a deviation
vector $|\Delta f\rangle = |f_1\rangle - |f_2\rangle $ and a
scalar product
\[
\langle \Delta f|\Delta f\rangle = \langle f_1|f_1 \rangle + \langle
f_2|f_2\rangle - \langle f_1|f_2\rangle - \langle f_2|f_1\rangle .
\]
Now we can define the relative deviation
\begin{equation}
\delta f = 1 - \frac{\langle f_1|f_2\rangle + \langle f_2|f_1\rangle}{\langle
f_1|f_1\rangle + \langle f_2|f_2\rangle}
\end{equation}
as a numerical estimate of the closeness of two functions. Note
that $\delta f = 0$ if $f_1(q) = f_2(q)$.

   Thus, we get the following estimate
\begin{equation}
\delta\Psi(n) = 1 - \langle \Psi_{ex}(n)|\Psi_{ap}(n)\rangle
\end{equation}
in the case of the normalized real functions $\Psi_{ex}(q,n)$ and
$\Psi_{ap}(q,n) .$ The same numerical comparison may be performed
\begin{equation}
\delta\Psi'(n) = 1
 - \frac{2 \langle \Psi'_{ex}(n)|\Psi'_{ap}(n)\rangle}
{\langle \Psi'_{ex}(n)|\Psi'_{ex}(n)\rangle  + \langle
\Psi'_{ap}(n)|\Psi'_{ap}(n)\rangle}
\end{equation}
for the first derivatives $\Psi'(q) = d\Psi(q)/d q $ . Naturally,
we can define analogous estimates for higher derivatives.

     Moreover, we estimate the closeness of functions $\hat H \Psi_{ap}(q,n)$
and $\hat H \Psi_{ex}(q,n)$ with the help of the following quantity
\begin{equation}
\delta \hat{H}\Psi(n) = \frac{\langle \Psi_{ap}(n)|\hat
H^2|\Psi_{ap}(n)\rangle + E_{ex}(n)^{2}( 1 - 2 \langle
\Psi_{ex}(n)|\Psi_{ap}(n)\rangle )} {\langle \Psi_{ap}(n)|\hat
H^2|\Psi_{ap}(n)\rangle + E_{ex}(n)^{2}}  ,
\end{equation}
where $E_{ex}(n)$ is the exact energy value.

     In addition, we compare two functions $\hat H \Psi_{ap}(q,n)$
and $\bar E(n) \Psi_{ap}(q,n)$. This comparison may be performed
when we do not know the exact solutions. As a result we get the
relative discrepancy
\begin{equation}
d(n) = \frac{\langle \Psi_{ap}(n)|\hat H^2|\Psi_{ap}(n)\rangle - \bar E(n)^2}
{\langle \Psi_{ap}(n)|\hat H^2|\Psi_{ap}(n)\rangle + \bar E(n)^2}
\end{equation}
which is directly connected with the Schr\"odinger equation under
consideration.

    Finally, we characterize our approximation by the usual relative
energy error
\begin{equation}
\triangle E(n) = \frac{\bar E(n)}{E_{ex}(n)} - 1 .
\end{equation}

    The verification is performed for three potentials with different
asymptotics. They are the harmonic oscillator potential
\[
V(q) = k q^2 ,
\]
the Morse potential
\[ V(q) = \frac{{\gamma}^2 {\hbar}^2 {\alpha}^2}{2 m}
 \left(e^{-2 \alpha q} - 2 e^{-\alpha q} \right)
\]
 and the modified P\"oschl-Teller potential
\[
  V(q) = \frac{\lambda (\lambda - 1) {\hbar}^2
{\alpha}^2} {2 m \cosh^2(\alpha q)} .
\]
The exact solutions of the Schr\"odinger equation for these
potentials may be found in \cite{land}.

\begin{center}
{\bf Table 4.1}  Numerical verification of the proposed
approximation.

\medskip
\begin{tabular}{r r r r r r}\hline
 $n$  &$\delta\Psi(n)$ & $\delta\Psi'(n)$ &$\delta \hat{H}\Psi(n)$& $ d(n)$ & $\triangle E(n)$
 \\ \hline
  \multicolumn{6}{c}{The harmonic oscillator potential.}  \\ \hline
 0 & $2.16 \cdot 10^{-4}$ & $4.60 \cdot 10^{-3}$ & $5.79 \cdot 10^{-2}$ &
  $5.32 \cdot 10^{-2}$ & $ 4.52 \cdot 10^{-3}$  \\
 1 & $1.59 \cdot 10^{-5}$ & $7.12 \cdot 10^{-5}$ & $1.33 \cdot 10^{-4}$ &
  $6.09 \cdot 10^{-5}$ & $ 5.64 \cdot 10^{-5}$  \\
 2 & $3.86 \cdot 10^{-6}$ & $1.05 \cdot 10^{-5}$ & $1.48 \cdot 10^{-5}$ &
  $3.78 \cdot 10^{-6}$ & $ 7.15 \cdot 10^{-6}$  \\
 3 & $1.50 \cdot 10^{-6}$ & $3.10 \cdot 10^{-6}$ & $4.12 \cdot 10^{-6}$ &
  $7.31 \cdot 10^{-7}$ & $ 1.89 \cdot 10^{-6}$  \\ \hline
 \multicolumn{6}{c}{The Morse potential  $\quad ( \gamma = 4.5 )$.} \\ \hline
  0 & $2.76 \cdot 10^{-4}$ & $5.98 \cdot 10^{-3}$ & $4.59 \cdot 10^{-3}$ &
   $5.77 \cdot 10^{-3}$ & $-1.45 \cdot 10^{-3}$   \\
  1 & $4.95 \cdot 10^{-5}$ & $2.34 \cdot 10^{-4}$ & $2.45 \cdot 10^{-4}$ &
   $3.56 \cdot 10^{-4}$ & $-1.60 \cdot 10^{-4}$   \\
  2 & $2.26 \cdot 10^{-5}$ & $8.22 \cdot 10^{-5}$ & $3.72 \cdot 10^{-4}$ &
   $4.51 \cdot 10^{-4}$ & $-1.01 \cdot 10^{-4}$   \\
  3 & $1.77 \cdot 10^{-5}$ & $6.43 \cdot 10^{-5}$ & $2.35 \cdot 10^{-3}$ &
   $2.49 \cdot 10^{-3}$ & $-1.59 \cdot 10^{-4}$   \\ \hline
 \multicolumn{6}{c}{The modified P\"oschl-Teller potential $\quad ( \lambda = 5 )$.}  \\ \hline
0 & $2.31 \cdot 10^{-4}$ & $6.60 \cdot 10^{-3}$ & $4.49 \cdot
10^{-3}$ &
   $5.70 \cdot 10^{-3}$ & $-1.44 \cdot 10^{-3}$   \\
  1 & $2.91 \cdot 10^{-5}$ & $2.10 \cdot 10^{-4}$ & $1.65 \cdot 10^{-4}$ &
   $2.97 \cdot 10^{-4}$ & $-1.62 \cdot 10^{-4}$   \\
  2 & $1.98 \cdot 10^{-5}$ & $6.17 \cdot 10^{-5}$ & $2.36 \cdot 10^{-4}$ &
   $3.18 \cdot 10^{-4}$ & $-1.01 \cdot 10^{-4}$   \\
  3 & $5.22 \cdot 10^{-5}$ & $4.90 \cdot 10^{-5}$ & $1.95 \cdot 10^{-3}$ &
   $2.05 \cdot 10^{-3}$ & $-1.48 \cdot 10^{-4}$  \\ \hline
\end{tabular}

\end{center}

    Table 4.1 shows that our approximation
gives fairly accurate results for all considered potentials and for all
considered quantities. Hence, the performed reconstruction of the WKB series
and subsequent explicit summation of the leading constituent (partial) series
yield the satisfactory (qualitative and quantitative) description of
 wave functions.


\begin{thebibliography}{99}
\bibitem{kudr1} Kudryashov V. V. 1997 {\it Quantum Systems: New Trends and
 Methods (Proceeding, Minsk, 1996)} ed Y. S. Kim {\it et al} (Singapore: World
  Scientific) pp. 202-205.
\bibitem{kudr2} Kudryashov V. V. 1998 {\it Doklady of the National Academy
 of Sciences of Belarus} {\bf 42} (no. 6) 45
(Russian).
\bibitem{boyd} Boyd J. P. 1999 {\it Acta Appl. Math.}
{\bf 56} 1.
\bibitem{bend1} Bender C. M. and Orszag S. A. 1978 {\it Advanced Mathematical
Methods for Scientists and Engineers} (New York: McGraw-Hill).
\bibitem{pham} Delabaere E., Dilinger H. and Pham F. 1997 {\it J. Math. Pys.} {\bf 38}
6126.
\bibitem{bend2} Bender C. M., Olaussen K. and Wang P. S. 1977 {\it Phys. Rev.} D {\bf 16}
1740.
\bibitem{robn}  Robnik M. and Romanovski V. G. 2000 {\it J. Phys. A: Math. Gen.} {\bf 33}
5093.
\bibitem{voros}  Voros A. 1999 {\it J. Phys. A: Math. Gen.} {\bf 32}
5993.
\bibitem{hand}  Abramovitz M. and Stegun I. A. 1970 {\it
 Handbook of Mathematical Functions} (New York: Dover Publications).
\bibitem{olver} Olver F. W. J. 1974 {\it Asymptotics and Special
Functions} (New York: Academic Press).
\bibitem{lang} Langer R. E. 1937 {\it Phys. Rev.} {\bf 51} 669.
\bibitem{land} Landau L. D. and Lifshitz E. M. 1977 {\it Quantum Mechanics}
 (Oxford: Pergamon Press).
\end{thebibliography}
\end{document}